\documentclass[final]{aipproc}
\layoutstyle{6x9}

\begin{document}

\title{Behavior of physical observables in the vicinity of the QCD critical end point}


\author{Pedro Costa}{address={Centro de F\'{\i}sica Te\'{o}rica,
Departamento de F\'{\i}sica, Universidade, P3004-516 Coimbra, Portugal}}





\begin{abstract}
Using the SU(3) Nambu-Jona-Lasinio (NJL) model, we study the chiral phase transition at finite $T$ and $\mu_B$. 
Special attention is given to the QCD critical end point (CEP): the study of physical quantities, as the pressure, the entropy, the baryon number susceptibility and the specific heat near the CEP, will provide complementary information concerning the order of the phase transition. We also analyze the information provided by the study of the critical exponents around the CEP. 
\end{abstract}

\maketitle


The existence of the CEP in QCD was suggested in the end of the eighties, and its properties have been studied since then (for a general review see Refs. \cite{Stephanov:2004,Casalbuoni:2006}).
The most recent lattice results for the study of dynamical QCD with $N_f=2+1$ staggered quarks of physical masses indicate the location of the CEP at $T^{CEP}=162\pm2\mbox{MeV},\,\mu^{CEP}=360\pm40\mbox{MeV}$ \cite{Fodor:2004JHEP}, however its exact location is not yet known once the location of the CEP depends strongly of the mass of the strange quark. At the CEP the phase transition is of second order, belonging to the three-dimensional Ising universality class, and this kind of phase transitions are characterized by long-wavelength fluctuations of the order parameter. 

As pointed out in \cite{Hatta:2003PRD,Schaefer:2006}, the critical region around the CEP is not pointlike but has a very rich structure. 
The vicinity of the CEP is a privileged region to study the influence of different type phase transitions in the physical observables, namely, the pressure, the entropy, the baryon number susceptibility, $\chi_B$, and the specific heat, $C_V$.

We perform our calculations in the framework of the  three--flavor  NJL   model, including the determinantal 't Hooft interaction that breaks the $U_A(1)$ symmetry, which has the  following Lagrangian: %
\begin{eqnarray} \label{lagr}
{\mathcal L} &=& \bar{q} \left( i \partial \cdot \gamma - \hat{m} \right) q
+ \frac{g_S}{2} \sum_{a=0}^{8}
\Bigl[ \left( \bar{q} \lambda^a q \right)^2+
\left( \bar{q} (i \gamma_5)\lambda^a q \right)^2
 \Bigr] \nonumber \\
&+& g_D \Bigl[ \mbox{det}\bigl[ \bar{q} (1+\gamma_5) q \bigr]
  +  \mbox{det}\bigl[ \bar{q} (1-\gamma_5) q \bigr]\Bigr] \, .
\end{eqnarray}
By using a standard hadronization procedure, an effective meson action is obtained, leading to the gap equations for the constituent quark masses from which several observables are calculated (we follow the methodology presented in detail in Refs. \cite{Costa:2003PRC,Costa:2002PLB}).

\begin{figure*}[t]
\resizebox{0.5\textwidth}{!}{%
  \includegraphics{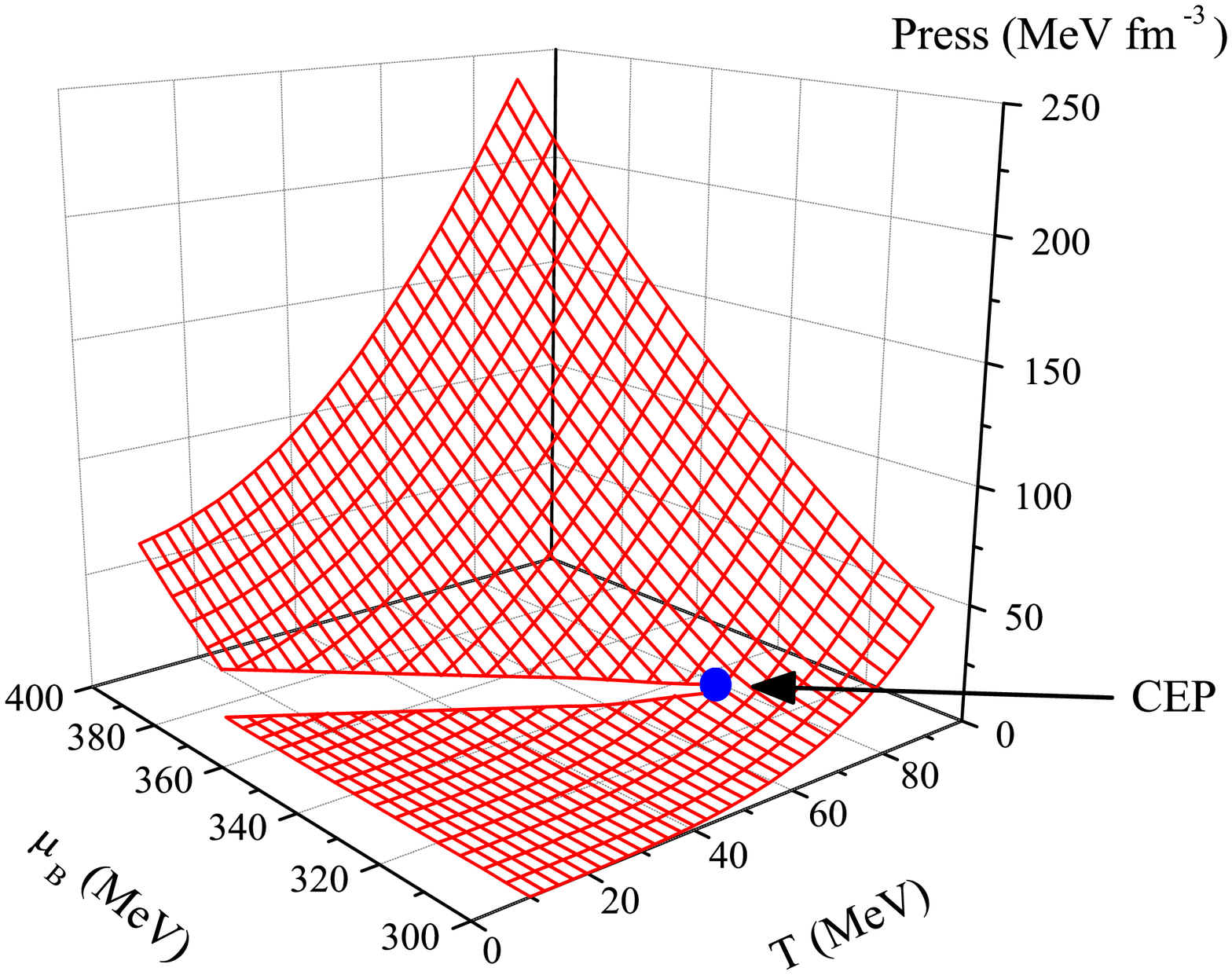}}
\resizebox{0.5\textwidth}{!}{%
  \includegraphics{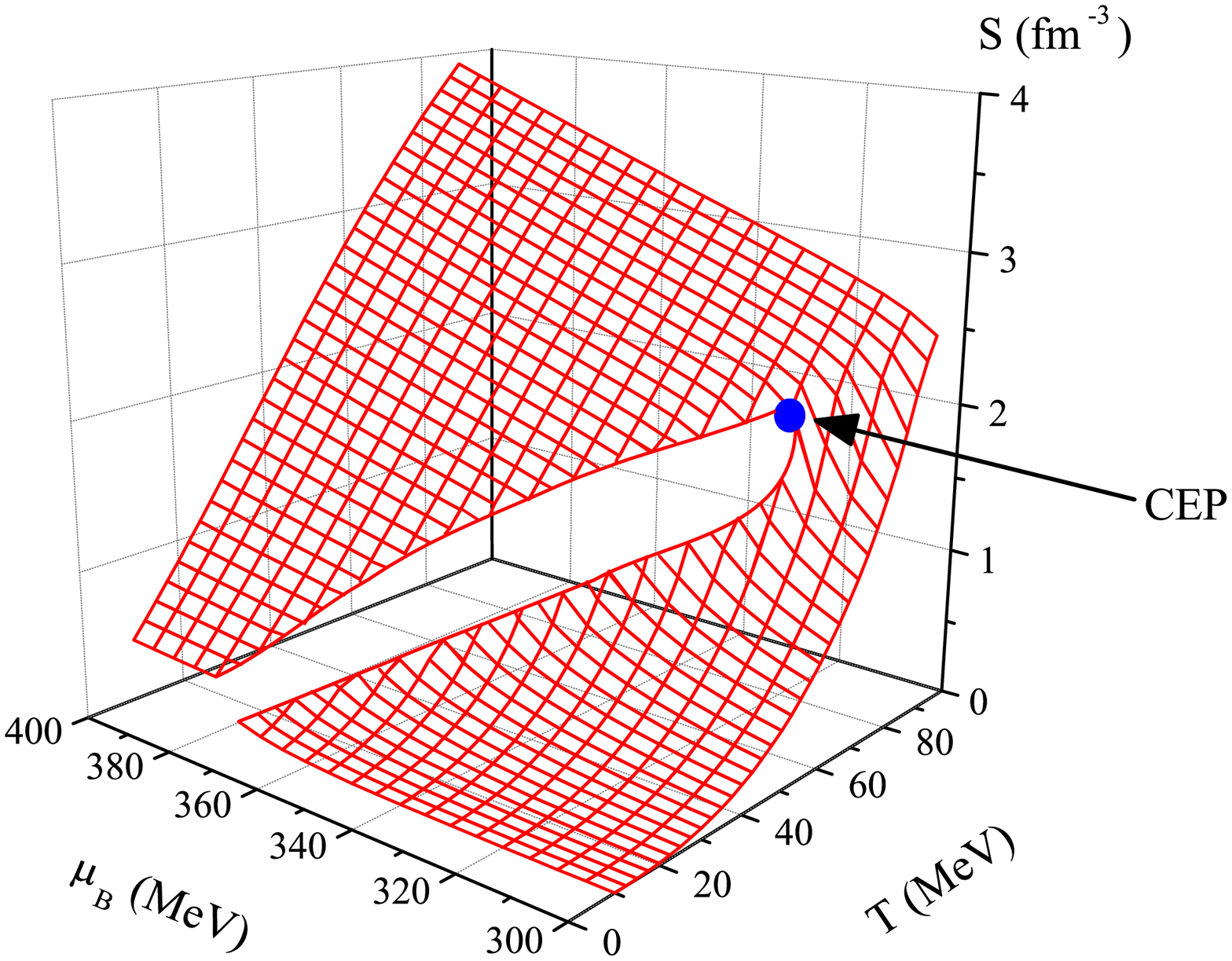}}
\caption{The pressure (left panel) and the entropy (right panel) as functions of the temperature and the baryonic chemical potential.}
\label{fig:1}       
\end{figure*}

The fundamental relation is provided by the baryonic thermodynamic potential
\begin{equation}\label{tpot}
	\Omega (\mu_i ,T)= E- TS - \sum_{i=u,d,s} \mu _{i} N_{i}\,,
\end{equation}
from which the relevant observables as the pressure $P$, the entropy $S$
and the particle number $N_i$ can be calculated as usually (the expressions are given in Ref. \cite{Costa:2003PRC}).

The nature of the chiral phase transition in NJL type models at finite $T$ and/ or $\mu_B$ has been discussed by different authors \cite{Costa:2003PRC}. 
For zero temperature the transition is of first order. As the temperature increases, the first order transition line persists up to the CEP. In the CEP the chiral transition becomes of second order. At higher temperatures there is a smooth crossover.

\begin{figure*}
\resizebox{0.5\textwidth}{!}{%
  \includegraphics{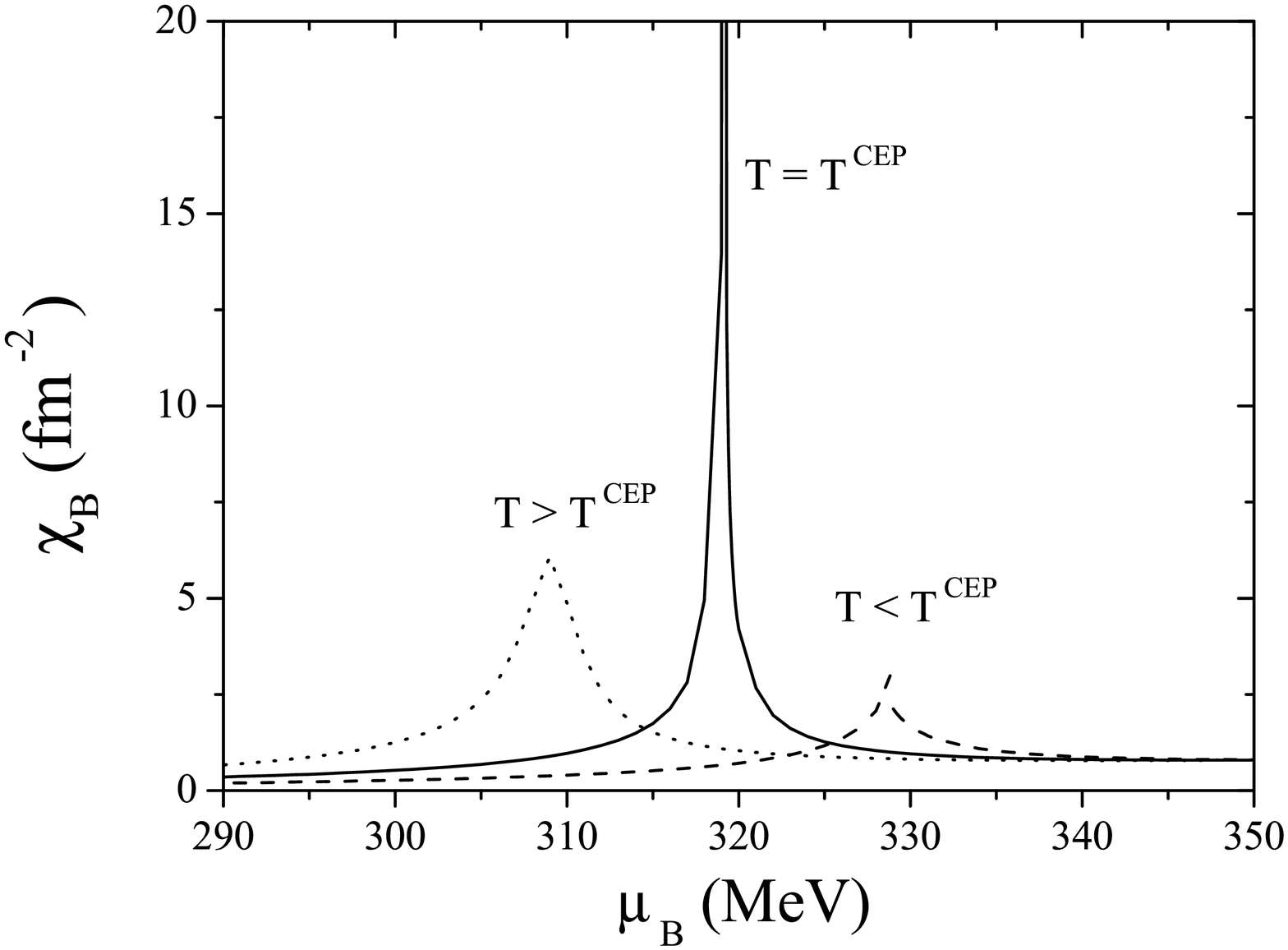}}
\resizebox{0.5\textwidth}{!}{%
  \includegraphics{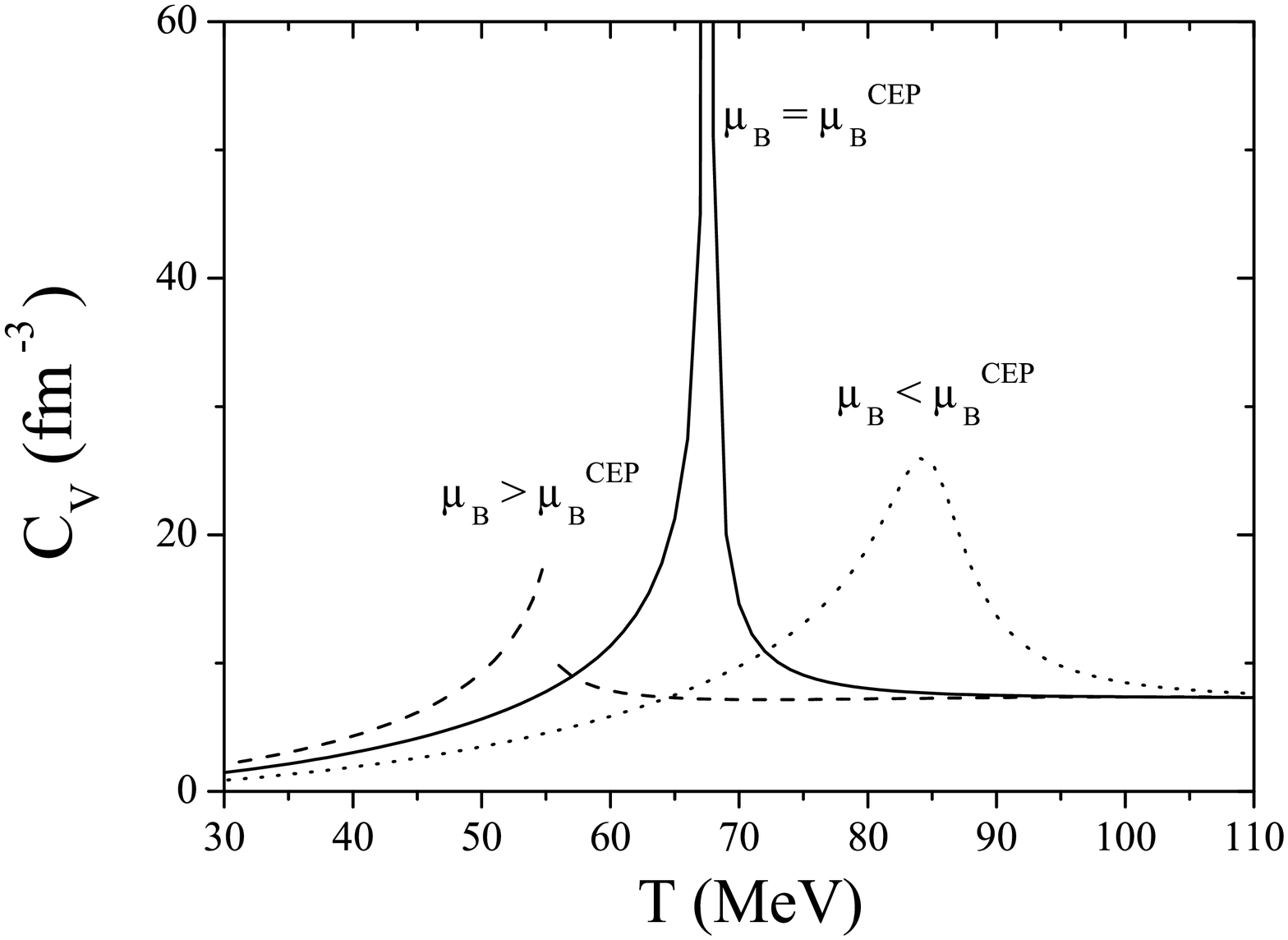}}
\caption{Left panel: baryon number susceptibility as a function of $\mu_B$ for different temperatures around the CEP: $T^{CEP}=67.7$ MeV and $T=T^{CEP}\pm10$ MeV. Right panel: specific heat as a function of $T$ for different values of $\mu_B$ around the CEP: $\mu_B^{CEP}=318.4$ MeV and $\mu_B=\mu_B^{CEP}\pm10$ MeV.}
\label{fig:2}       
\end{figure*}

In the vicinity of the CEP, the nature of the chiral phase transition influence strongly the behaviour of the physical observables. In the first order phase transition region ($T<T^{CEP},\,\mu_B>\mu_B^{CEP}$) both, pressure and entropy, show a discontinuity which end at the CEP. In the crossover region ($T>T^{CEP},\,\mu_B<\mu_B^{CEP}$), the discontinuities of $P$ and $S$ vanish, and both quantities change gradually in a continuous way (see Fig. \ref{fig:1}).
The same situation can be seen in the behaviour of $\chi_B$: in the first order phase transition region $\chi_B$ has a discontinuity (left panel of Fig. \ref{fig:2}). At the CEP, the phase transition is of second order, and the slope of the baryonic density tends to infinity which implies a diverging $\chi_B$. In the crossover region, $\chi_B$ changes gradually in a continuous way. 
A similar behavior is found for the specific heat for three different chemical potentials around the CEP, as we can observe from the right panel of Fig. \ref{fig:2}.

Focusing our attention on the critical behavior of the baryon number susceptibility in the vicinity of the CEP, we verify that $\chi_B$ diverges with a certain critical exponent. 
Considering a path parallel to the $\mu_B$-axis in the ($T,\mu_B$)-plane from lower $\mu_B$ towards the critical $\mu_B^{CEP}= 318.4$ MeV at fixed temperature $T^{CEP} = 67.7$ MeV, one can calculate the critical exponent $\epsilon$ using the linear logarithmic fit $\ln \chi_B = -\epsilon \ln |\mu_B -\mu_B^{CEP} | + const$, where the term $const$ is independent of $\mu_B$. The result that we obtain is $\epsilon = 0.67\pm 0.01$, which is consistent with the mean field theory prediction: $\epsilon = 2/3$. 
Since there is no reason why the critical exponent should be equal for both regions, below and above $\mu_B^{CEP}$, we also study $\chi_B$ from higher $\mu_B$ towards the critical $\mu_B^{CEP}$. Using again a logarithmic fit, the result is $\epsilon' = 0.68\pm 0.01$ which is very near the value of $\epsilon$. This means that the size of the region we observe is approximately the same independently of the direction we choose in the path parallel to the $\mu_B$-axis. 
Complementary information is also obtained from the study of the critical exponent of $C_V$ \cite{Costa:2006}.

Summarizing our discussion, we have analyzed the vicinity of the QCD critical end point in the SU(3) NJL model.
We conclude that, in this region, the physical observables are strongly influenced by the nature of the phase transition. 
Around the CEP we have studied the baryon number susceptibility which is related with event-by-event fluctuations of $\mu$ in heavy-ion collisions. The study of the specific heat is also important once it is related with event-by-event fluctuations of $T$  in heavy-ion collisions \cite{Stephanov:1998PRL}. 
We also conclude that the critical exponents of $\chi_B$ obtained in our model are consistent with the mean field value $\epsilon\simeq\epsilon'\simeq2/3$. 

\vspace{0.5cm}
Work supported by grant SFRH/BPD/23252/2005 from F.C.T. and Centro de F\'{\i}sica Te\'orica.




\begin{thebibliography}{9}

\bibitem{Stephanov:2004}
		M. A. Stephanov, 
		Prog. Theor. Phys. Suppl. 153 (2004) 139; 
		Int. J. Mod. Phys. A 20 (2005) 4387.
		
\bibitem{Casalbuoni:2006}    
		R. Casalbuoni,
    hep-ph/0610179.

\bibitem{Fodor:2004JHEP} 
		{Z. Fodor, S. D. Katz}, 
		J. High Energy Phys. 0204 (2004) 050. 

\bibitem{Hatta:2003PRD}		
		Y. Hatta, T. Ikeda,		
		Phys. Rev. D 67 (2003) 014028.

\bibitem{Schaefer:2006}
		B.-J. Schaefer, J. Wambach, hep-ph/0603256. 

\bibitem{Costa:2003PRC}
		P. Costa, M. C. Ruivo, Y. L. Kalinovsky, C. A. de Sousa, 
 		Phys. Rev. C 70 (2004) 025204.

\bibitem{Costa:2002PLB} 
		P. Costa, M. C. Ruivo, Yu. L. Kalinovsky, 
		Phys. Lett. B 560 (2003) 171.

\bibitem{Costa:2006}
		P. Costa, C. A. de Sousa, M. C. Ruivo, Y. L. Kalinovsky, 
		hep-ph/0701135.

\bibitem{Stephanov:1998PRL}
		M. Stephanov, K. Rajagopal, E. Shuryak,
		Phys. Rev. Lett. 81 (1998) 4816.

\end{thebibliography}
\end{document}